\documentclass{aa}  
\usepackage{natbib}
\bibpunct{(}{)}{;}{a}{}{,} % to follow the A&A style
\usepackage[normalem]{ulem}
\usepackage{xcolor}
\usepackage{siunitx}

\usepackage{graphicx}
\usepackage{txfonts}
\begin{document} 

   \title{Even redder than we knew: color and $A_{\mathrm{V}}$ evolution up to $z=2.5$ from JWST/NIRCam Photometry}
   \titlerunning{Colors and $A_{\mathrm{V}}$}

   \author{A.~van der Wel\inst{1} \and
          M.~Martorano\inst{1} \and
          D.~Marchesini\inst{2} \and
          S.~Wuyts\inst{3} \and
          E.~F.~Bell\inst{4} \and
          S.~E.~Meidt\inst{1} \and
          A.~Gebek\inst{1} \and
          G.~Brammer\inst{5} \and
          K.~Whitaker\inst{5,6} \and
          R.~Bezanson\inst{7} \and
          E.~J.~Nelson\inst{8} \and
          G.~Rudnick\inst{9} \and
          M.~Kriek\inst{10} \and
          J.~Leja\inst{11,12,13} \and
          J.~S.~Dunlop\inst{14} \and
          C.~Casey\inst{15,16,5} \and
          J.~Kartaltepe\inst{17}
          }
    %\orcidlink{0000-0002-1404-5950}

   \institute{Sterrenkundig Observatorium, Universiteit Gent, Krijgslaan 281 S9, 9000 Gent, Belgium\\
              \email{arjen.vanderwel@ugent.be}
         \and
        Physics and Astronomy Department, Tufts University, 574 Boston Avenue, Medford, MA 02155, USA
        \and
        Department of Physics, University of Bath, Claverton Down, Bath BA2 7AY, UK
        \and
        Department of Astronomy, University of Michigan, 1085 South University Avenue, Ann Arbor, MI, 48109–1107, USA
        \and
        Cosmic Dawn Center (DAWN), Niels Bohr Institute, University of Copenhagen, Jagtvej 128, København N, Copenhagen DK-2200, Denmark
        \and
        Department of Astronomy, University of Massachusetts, Amherst, MA 01003, USA      
        \and
        Department of Physics and Astronomy and PITT PACC, University of Pittsburgh, Pittsburgh, PA 15260, USA
        \and
        Department for Astrophysical and Planetary Science, University of Colorado, Boulder, CO, 80309, USA
        \and
        University of Kansas, Department of Physics and Astronomy, 1251 Wescoe Hall Drive, Room 1082, Lawrence, KS 66049, USA
        \and
        Leiden Observatory, Leiden University, P.O. Box 9513,NL-2300 AA Leiden, The Netherlands
        \and
        Department of Astronomy \& Astrophysics, The Pennsylvania State University, University Park, PA 16802, USA
        \and
        Institute for Computational \& Data Sciences, The Pennsylvania State University, University Park, PA 16802, USA
        \and
        Institute for Gravitation and the Cosmos, The Pennsylvania State University, University Park, PA 16802, USA
         \and
             Institute for Astronomy, University of Edinburgh, Royal Observatory, Edinburgh, EH9 3HJ, UK
        \and
        Department of Physics, University of California, Santa Barbara, Santa Barbara, CA 93109, USA
        \and
        Department of Astronomy, The University of Texas at Austin, Austin, TX, USA
        \and
        Laboratory for Multiwavelength Astrophysics, School of Physics and Astronomy, Rochester Institute of Technology, 84 Lomb Memorial Drive, Rochester, NY 14623, USA
             }

   \date{Received April 10, 2025; accepted June 30, 2025}

% \abstract{}{}{}{}{} 
% 5 {} token are mandatory
 
  \abstract
   {}
   {JWST/NIRCam provides rest-frame near-IR photometry of galaxies up to $z=2.5$ with exquisite depth and accuracy. This affords an unprecedented view of the evolution of the UV-optical-near-IR color distribution and its interpretation in terms of the evolving dust attenuation, $A_{\mathrm{V}}$.}
   {
   We use the value-added data products (photometric redshift, stellar mass, rest-frame $U-V$ and $V-J$ colors, and $A_{\rm V}$) provided by the public DAWN JWST Archive. These data product derive from fitting the spectral energy distributions obtained from multiple NIRCam imaging surveys, augmented with pre-existing HST imaging data.
   Our sample consists of a stellar mass complete sample of $\approx 28,000$ $M_\star> 10^{9}~M_\odot$ galaxies in the redshift range $0.5<z<2.5$.}
   {The $V-J$ color distribution of star-forming galaxies evolves strongly, in particular for high-mass galaxies ($M_\star>3\times 10^{10}~M_\odot$), which have a pronounced tail of very red galaxies reaching $V-J> 2.5$ at $z>1.5$ that does not exist at $z<1$. Such red $V-J$ can only be explained by dust attenuation, with typical values for $M_\star \approx 10^{11}~M_\odot$ galaxies in the range $A_{\mathrm{V}}\approx 1.5-3.5$ at $z\approx 2$.
   This redshift evolution went largely unnoticed before because the photometric redshift estimates for the reddest ($V-J>2.5$), most attenuated galaxies has markedly improved thanks to the new, precise photometry, in much better agreement with the 25 available spectroscopic redshifts for such galaxies.
   %, frequently reaching the 16-to-84-percentile range of attenuation values increases from $A_{\mathrm{V}}\approx 0.8-2.1$ at $z<1$ to $A_{\mathrm{V}}\approx 1-3.5$ at $2<z<2.5$. 
   The reddest population readily stands out as the independently identified population of galaxies detected at sub-mm wavelengths.
   Despite the increased attenuation, $U-V$ colors across the entire mass range are slightly bluer at higher $z$. A well-defined and tight color sequence exists at all redshifts $0.5<z<2.5$ for $M_\star>3\times 10^{10}~M_\odot$ quiescent galaxies, in both $U-V$ and $V-J$, but in $V-J$ it is blue rather than red compared to star-forming galaxies.
   In conclusion, whereas the rest-frame UV-optical color distribution evolves remarkably little from $z=0.5$ to $z=2.5$, the rest-frame optical-near-IR color distribution evolves strongly, primarily due to a very substantial increase with redshift in dust attenuation for massive galaxies.}
    {}

   \maketitle
%
%
%-------------------------------------------------------------------

\section{Introduction}\label{sec:Introduction}
UV-optical color-magnitude and color-mass diagrams of galaxies have long been used as a descriptive tool to understand the distribution of galaxy properties and their evolution with cosmic time. \citet{holmberg50} was the first to note the overall correlation between color and galaxy type.
\citet{pettit54, baum59, de-vaucouleurs61} showed a correlation between color and luminosity.
These fundamental results serve as a precursor of the concept "red sequence", a term coined by \citet{gladders98}, who were inspired by the key insight that the optical color-magnitude relation for early-type galaxies is universal and extremely tight, with $<$0.1 mag scatter \citep{bower92}. A red sequence was also found, with similarly small scatter, in clusters at large look-back time \citep[$z\lesssim 1$,][]{ellis97, stanford98, gladders98}, a discovery made possible by the then newly available precise, stable and deep photometry afforded by the Hubble Space Telescope. 

Up until that point, determining the evolution of the red sequence mostly focused on massive ellipticals and galaxy clusters and the mainstream view was that red galaxies constitute a passively evolving population that formed at very early cosmic times. But deep, wide-area surveys would soon lead to a paradigm shift. The terminology -- the red sequence and its counterpart, the "blue cloud" \citep[first mentioned in][]{phleps06} -- was used in seminal studies in the mid-2000s to effectively communicate the key new insight that since at least $z\approx 1$ there has been a population of galaxies with low star-formation activity -- for which the small range in color is set by the slowly evolving colors of old stellar population -- and a more actively star-forming population, the colors of which vary by larger amounts due to young stars and dust attenuation by dust \citep{blanton03a, bell04a, bell04, faber07, brown07}.  Arguably the most important aspect of the paradigm shift is that the number density of red sequence galaxies grew by a factor $\approx 2$ since $z=1$ \citep{bell04a}. This population does not evolve passively. Many of the progenitors of present-day red-sequence galaxies must have been actively forming stars at $z\lesssim 1$ (hence, the term "progenitor bias", \citealt{van-dokkum96}) and, in the meantime, joined the red sequence by ceasing this star-forming activity through some process or combination of processes now commonly referred to under the umbrella of quenching \citep[as in, e.g.,][]{faber07}.\footnote{The term "quenching" in reference to galaxy-wide cessation of star formation caused by a heating or feedback process was, to the best of our knowledge, first used by \citet{cox83}. This paper was cited once in the past decade, and only in the context of molecular cloud properties.} Add to this the discovery of substantial structural evolution of red galaxies \citep[e.g.,][]{van-der-wel08b, van-dokkum08a} and the paradigm shift is complete: there is no passive evolution of either the population or the individual members.

Evidence for the existence of a red sequence at earlier cosmic times, beyond $z=1$, emerged gradually, first through the discovery of a population of red galaxies \citep{dickinson00, smail02a, franx03} at $z\approx 2$, which were shown to be mix of evolved and dusty galaxies \citep[e.g.,][]{smail02a, dunlop07, wuyts07, arnouts07, kriek08, Brammer09}. Since then, it has become customary to characterize the galaxy population and its evolution in terms of a star-forming (main) sequence \citep{noeske07} and a quiescent population, either by separating the sub-groups on the basis of their specific star-formation rate \citep[e.g.,][]{whitaker14} or in a color-color diagram, such as $U-V$ vs.~$V-J$. These approaches allow to differentiate between old stellar populations and younger, dust-reddened galaxies \citep[e.g.,][]{labbe05, williams09, brammer11, whitaker11, muzzin13, tacchella18, leja19c}.  

The goal of this short paper is to revisit the redshift evolution of the color distribution of galaxies. The immediate motivation is that with JWST/NIRCam we can now measure very accurate and precise rest-frame optical-to-near-infrared colors for the general galaxy population up to $z\approx 2.5$. We will demonstrate significant evolution in the rest-frame $V-J$ color and a strong increase in dust attenuation by dust for massive star-forming galaxies over from $z=0.5$ to $z=2.5$.

The improved insights into dusty galaxies at $z\gtrsim 2$ must also be seen in the context of parallel developments in far-IR and (sub-)mm astronomy. A class of sub-mm galaxies at $z\approx 2$, highly obscured in the optical and very luminous in thermal radiation from dust, was discovered in the late 90s \citep{smail97, hughes98}, and has been an active area of research since \citep[e.g.,][]{smail02a, chapman03, daddi07, casey13, dunlop17, mclure18, tacconi18, adscheid24}. Likewise, red, high-$z$ galaxies connect to (ultra-)luminous infrared galaxies, first discovered \citep{soifer84} in the present-day Universe as exceptional, merger-driven starbursts \citep{lonsdale84, armus87}, and later at higher redshift, as relatively common occurrences \citep[e.g.,][]{lonsdale04}, reflective of two-orders of magnitude higher star-formation levels overall than ordinary star-forming galaxies. Without attempting to systematically connect the reddest galaxies found with NIRCam to the populations identified at longer wavelengths, we will provide the necessary context by cross-matching our sample with the ALMA sub-mm selected compilation from \citet{adscheid24} and show this population overlaps and connects with our stellar mass-selected sample.

We assume a standard, flat $\Lambda$CDM cosmology with H$_0=70~$km$~$s$^{-1}~$Mpc$^{-1}$ and $\Omega_m=0.3$, we adopt the AB magnitude system \citep{oke83}, and the \citet{chabrier03} IMF.

        \begin{figure*}[ht]
                \centering
        	\includegraphics[scale=0.44]{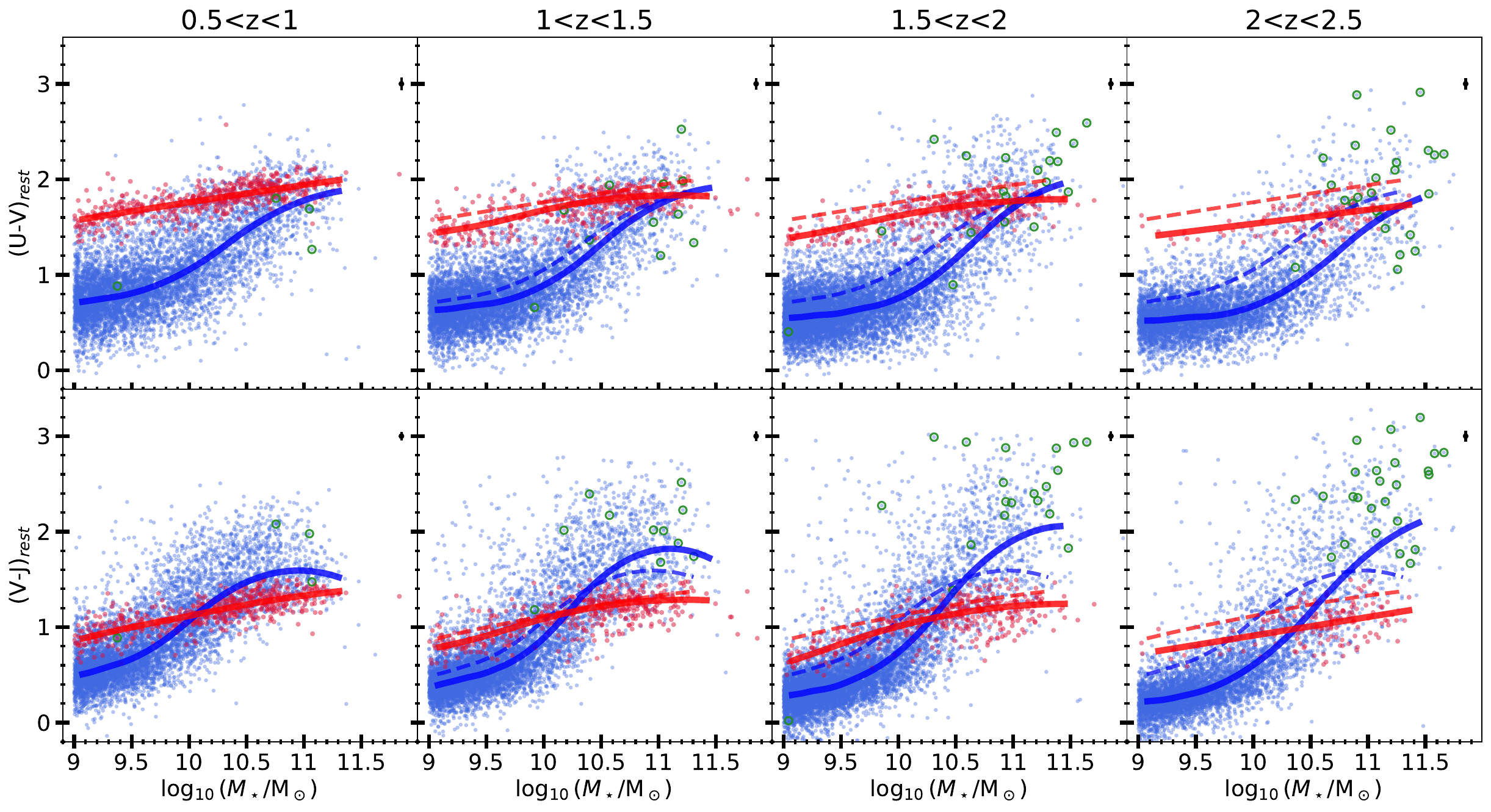}
                \caption{Rest-frame U-V (upper panels) and V-J (lower panels) colors versus stellar mass in four redshift bins. Star-forming galaxies are shown in blue; quiescent in red, as separated  by their location in the $V-J$ vs.~$U-V$ color-color diagram as defined by \cite{muzzin13} and shown here in Fig.~\ref{fig:uvj}.
                Solid lines are running medians, calculated with a spline-quantile regression \citep[COBS library][]{ng07,ng22}. The dashed lines show the running median lines from the $0.5<z<1$ (left-most) panels. Green circles indicate sub-mm-detected galaxies. 
                The error bars in the top right corner of each panel show the median random uncertainty on the rest-frame colors, including the measurement uncertainty of the observed fluxes and the propagated uncertainties from the SED fit.}
            \label{fig:Global_UV-VJ_zbins_ccAv}
        \end{figure*}

        \begin{figure*}[ht]
                \centering
        	\includegraphics[scale=0.4]{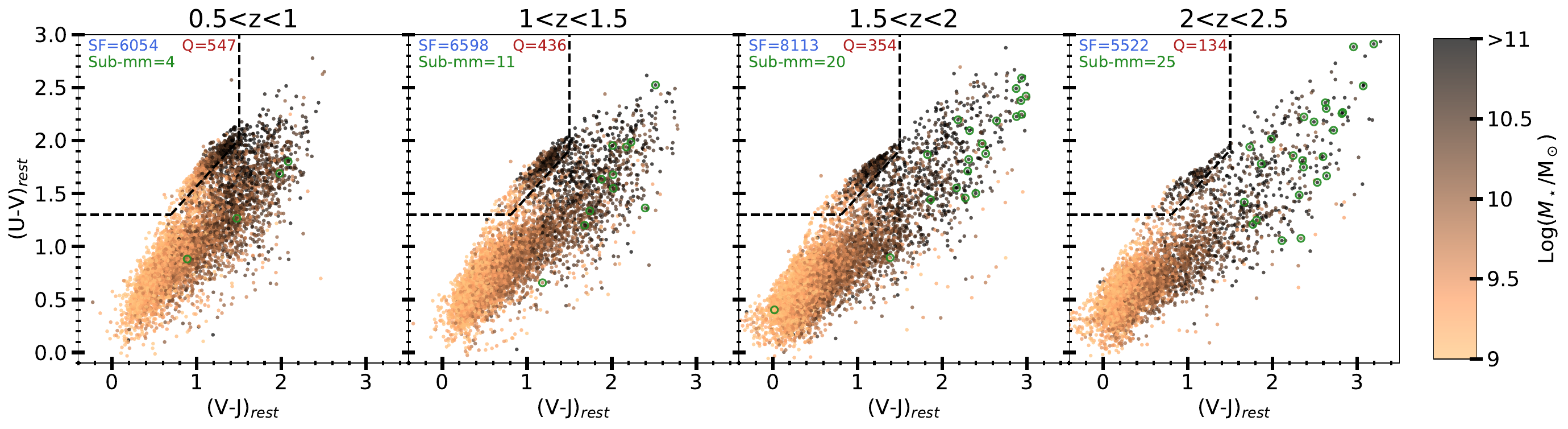}
                \caption{Rest-frame U-V vs.~V-J diagram in four redshift bins, showcasing the rise of very red galaxies at higher $z$ and indicating the separation between star-forming and quiescent galaxies.}
            \label{fig:uvj}
        \end{figure*}

\section{Data and sample selection}\label{sec:Data}
    %\subsection{Initial sample selection}\label{sec:sample}

    The parent sample is drawn from the Dawn JWST Archive (DJA) catalog\footnote{\url{https://dawn-cph.github.io/dja/blog/2024/08/16/morphological-data/}}. This catalog contains photometry for over $400,000$ galaxies in the five CANDELS \citep{grogin11, koekemoer11} fields observed with JWST/NIRCam as part of CEERS \citep{finkelstein23, finkelstein25}, PRIMER \citep{dunlop21}, COSMOS-Web \citep{casey23}, and JADES \citep{eisenstein23}.

    The data reduction process, the photometry and the SED fitting approach are described by \cite{valentino23}. In short, we use their \SI{0.5}{\arcsecond} diameter aperture photometry on all available HST/ACS, HST/WFC3,  and JWST/NIRCam imaging, ranging from $0.4\mu{\text{m}}$ to $4.4\mu{\text{m}}$. Two aperture corrections then provide total flux densities. The first correction accounts for the fraction of the flux for a point source outside the \SI{0.5}{\arcsecond} aperture. This correction varies from filter to filter. The second correction is the multiplication by the ratio the \textit{AUTO} flux and the aperture flux as measured in the detection image (typically, F277W+F356W+F444W). This correction is the same for all filters and accounts for the spatial extent of the source, but not color gradients. Colors gradients can be significant, but we verified that the colors within two F444W effective radii from the S\'ersic models presented by \citet{martorano25} are not systematically different from our aperture colors. We prefer the aperture colors for this work because of their small uncertainties and model independence.
    
    Photometric redshifts are inferred with \textsc{EAZY} \citep{brammer08},
    with the \textit{agn$\_$blue$\_$sfhz$\_$13} set of templates \footnote{\url{https://github.com/gbrammer/eazy-photoz/tree/master/templates/sfhz}}, which consists of 13 templates from Flexible Stellar Populations Synthesis \citep[FSPS][]{conroy09,conroy10}, a template constructed from the NIRSpec spectrum of an extreme emission-line galaxy at $z=8.5$, and a template generated to match the JWST/NIRSpec spectrum of a $z=4.5$ source which is thought to host an obscured AGN \citep{killi24}. As input all available HST and JWST imaging datasets were used. For a detailed, field-by-field list we refer to Table 3 in \citet{valentino23}, but for every object at least 12 wide filter photometric data points across the full wavelength range are available.

    We select galaxies with stellar mass $M_\star>10^{9}~M_\odot$, which are all at least a magnitude brighter than the 5$\sigma$ detection limit of PRIMER, the shallowest imaging dataset \citep{valentino23}, ensuring a stellar-mass complete sample.
    We limit the redshift range to $0.5<z<2.5$ to avoid overly large aperture corrections at low redshift and the loss of rest-frame near-IR ($J$-band) coverage at high redshift. We also reject 3,708 galaxies with poor SED fits ($\chi^2>10$).
    Our final sample consists of 27,758 galaxies, with as key parameters for this work: $z_{\mathrm{ph}}$, $M_\star$, rest-frame $U-V$, rest-frame $V-J$ and $A_{\mathrm{V}}$.

%
%
%--------------------------------------------------------------------
\section{Results} \label{sec:global_col}
\subsection{Evolution and stellar mass dependence of rest-frame colors} \label{sec:global_col}
        
        \begin{figure*}
                \centering
        	\includegraphics[scale=0.39]{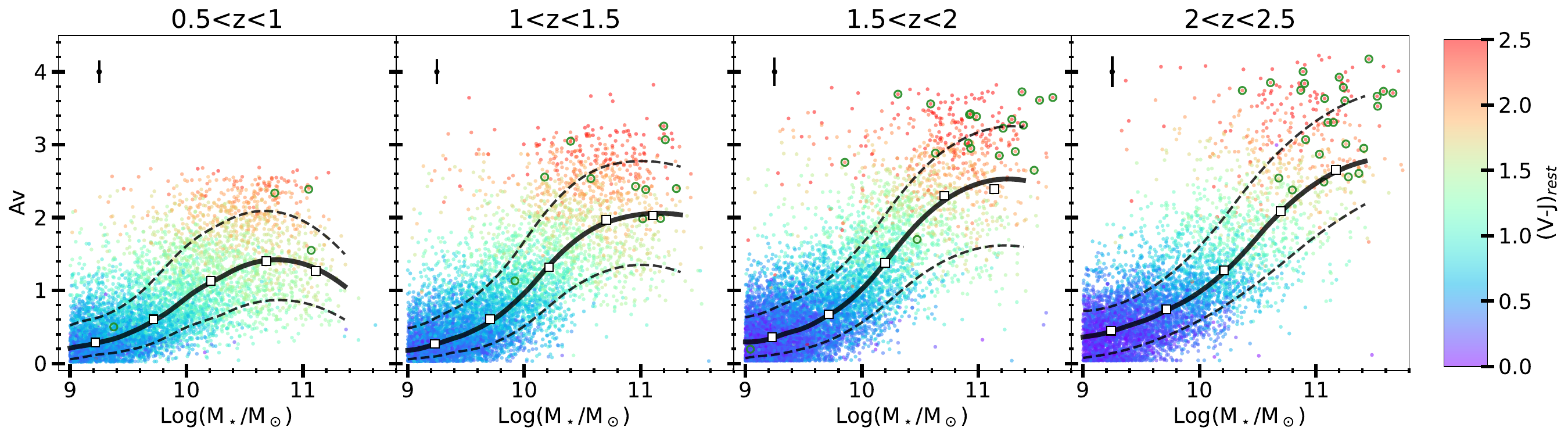}
                \caption{$A_{\mathrm{V}}$ vs.~$M_\star$ for star-forming galaxies in four redshift bins. Points are color-coded with $V-J$ color. White squares show the median in stellar mass bins and errorbars the statistical uncertainty ($\sigma/\sqrt{N}$, mostly smaller than the data points themselves). The solid black lines show the spline-percentile regression while the dashed lines show the 16-84 percentiles of the distribution. The error bars are the median uncertainties from the observed fluxes and propagated uncertainties from the SED fit.}
            \label{fig:AV_mass_zbins_ccVJ}
        \end{figure*}
        
        The evolution of the color-$M_\star$ distribution is shown in Figure \ref{fig:Global_UV-VJ_zbins_ccAv}.
        The redshift evolution in the $(V-J)$-$M_\star$ distribution for star-forming galaxies is quite remarkable. The $(V-J)$-$M_\star$ relation is very steep at $z>2$, with a nearly 2 mag difference between $M_\star\gtrsim 10^{11}~M_\odot$ and $M_\star < 10^{10}~M_\odot$ galaxies. Low-mass galaxies become bluer with redshift (by about 0.3 mag from $z<1$ to $z>2$), indicative of higher levels of star-formation activity, without strong attenuation.  High-mass galaxies, on the other hand, become redder with redshift. 
        
        Such red colors ($V-J\gtrsim 2$) can only be plausibly explained by dust reddening, since quiescent galaxies, which presumably have the oldest light-weighted ages, are not redder than $V-J\approx 1.2$. This is the reason for the popularity of color-color diagrams, shown here in Fig.~\ref{fig:uvj}, to separate dusty galaxies from quiescent galaxies: dusty galaxies are redder in $V-J$ \citep{williams09, ilbert09, whitaker11} than even the oldest stellar populations, which are found in quiescent galaxies according to state-of-the-art spectroscopic data and modeling \citep{gallazzi14, chauke18, kaushal24, nersesian25}. 
        These dusty, massive galaxies have been seen in near-IR surveys for decades, but in most previous works the colors of the reddest objects remained underestimated. The tail of red galaxies in $V-J$ did not, generally, extend much beyond $V-J\approx 2$ \citep[e.g.,][]{muzzin13, van-der-wel14, fang18}. Only \citet{martis19} showed a tail of galaxies at $V-J\approx 2.5$. With the new NIRCam photometry and current data analysis we see that the average color is $V-J\approx 2$ for massive star-forming galaxies, with a tail reaching $V-J\approx 3$. The issue is not that the reddest galaxies were undetected in HST surveys, but that their redshifts were generally overestimated and their rest-frame colors uncertain. Specifically, cross-matching the \textsc{ASTRODEEP-JWST} spectroscopic redshift ($z_{\rm sp}$) catalog \citep{merlin24} with the DJA catalog used in this paper and the 3D-HST catalog \citep{skelton14, momcheva16}, we find 25 galaxies with measured $z_{\rm sp}$ and $V-J>2.5$. Our median $z_{\mathrm{ph}}=1.64$ matches well with the median $z_{\mathrm{sp}}=1.68$, while in the 3D-HST catalog the median is $z_{\mathrm{ph, 3DHST}}=2.12$. For the full sample of 7124 matched objects there is no such offset: both sets of $z_{\rm ph}$ estimates agree very well, to $\approx$1\% on average, with $<1\%$ catastrophic outliers (defined as deviating by more than 15\% in $1+z$), with $\approx 5\%$ scatter, and with no systematic departures up to $V-J=2.5$. Even galaxies with with $2<V-J<2.5$ show no systematic offsets, which are apparently limited to the small subset of the very reddest objects which nonetheless produce a pronounced change in the color distribution of $z\approx 2$  galaxies when compared to lower $z$. It remains to be determined whether the new $z_{\rm ph}$ estimates are improved because of improved photometry or because of improved templates. 

        As mentioned in the Introduction, the discovery of a population of dusty galaxies at $z\gtrsim 2$ in near-IR surveys was made in parallel with observations at longer wavelengths \citep[e.g.,][]{hughes98, smail97, smail02, chapman03, daddi07}. 
        To provide this context, we cross-match our sample with the \textit{A$^3$COSMOS} and \textit{A$^3$GOODSS}  catalogs \citep{adscheid24} that collect all objects identified by ALMA across the COSMOS and GOODS fields. We find a match for 60 galaxies with separation $<$\SI{0.4}{\arcsecond}, all but one of which satisfy the homogeneous peak $S/N=5.40$ criterion for blind extraction.  The choice for $<$\SI{0.4}{\arcsecond} serves as a compromise that allows for spatial offsets between stellar and dust emission while avoiding spurious matches. Reducing or increasing the search radius by \SI{0.1}{\arcsecond} changes the number of matches by just three. This sub-mm detected sample is not complete or representative for our sample, and it should be borne in mind that essentially all massive, star-forming galaxies are sufficiently bright at sub-mm wavelengths to be detected in sufficiently deep ALMA observations \citep{dunlop17, mclure18}.
        
        As shown in Fig.~\ref{fig:Global_UV-VJ_zbins_ccAv}, the sub-mm-detected galaxies have typical stellar masses of $M_{\star}=10^{11}~M_\odot$ and are among the reddest 50\% of the population in $V-J$. But they have average $U-V$ colors, which illustrates that $U-V$ color is a poor indicator of attenuation and far-infrared luminosity. This is also illustrated by the strong evolution in $V-J$ for massive star-forming galaxies --- indicative of strong evolution in dust attenuation -- in the absence of a similarly strong evolution in $U-V$. The $U-V$ color, it seems, is a random draw, independent of dust mass and the level of star-formation. Such a random draw can be explained by patchiness of dust in the UV, with a small number of random lines of sight that happen to reveal a small fraction of the young stellar population \citep{zhang23, gebek25}. 

        A clearly defined, tight color sequence exists for quiescent galaxies in both $U-V$ and $V-J$, but at all redshifts $V-J$ is blue rather than red compared to equally massive star-forming galaxies: the red sequence in $U-V$ becomes a blue sequence in $V-J$. Quiescent galaxies become bluer with redshift in both $U-V$ and $V-J$ due to younger ages at earlier cosmic times. Similarly, the $U-V$ colors of star-forming galaxies are bluer at higher redshifts \citep[also see, e.g.,][]{marchesini14}. These trends are, of course, well documented and expressed in terms of a blue cloud of low-mass star-forming galaxies and a red sequence of quiescent galaxies that dominate the red galaxies at $z\lesssim 1$ \citep{bell04a, faber07}, with a minority of dust-reddened, star-forming galaxies mixed in at high mass \citep{bell04}. But beyond $z=1$ the quiescent red sequence in a color-mass or color-luminosity diagram becomes overwhelmed by dusty star-forming galaxies \citep[e.g.,][]{wuyts07, arnouts07} that are, on average, equally red but have a larger scatter in $U-V$. 
        
\subsection{Evolution and stellar mass dependence of $A_{\mathrm{V}}$ for star-forming galaxies} \label{sec:Av}
        The connection between color and dust attenuation for star-forming galaxies\footnote{$A_{\mathrm{V}}$ estimates of quiescent galaxies carry the additional uncertainty due to our limited knowledge of the intrinsic near-IR colors of evolved stellar populations. There is compelling evidence of somewhat dusty quiescent galaxies \citep[e.g.,][]{siegel25}, but it is a distinct possibility that SED fitting codes use reddening to correct for template mismatches \citep[e.g., Appendix B in][]{van-der-wel21}.} is made explicit in Figure \ref{fig:AV_mass_zbins_ccVJ}. The most striking feature is the substantial increase in $A_{\mathrm{V}}$ with redshift for massive star-forming galaxies. $A_{\mathrm{V}}\approx 1.2$ for $M_\star\gtrsim 10^{11}~M_\odot$ galaxies at $z<1$, increasing to $A_{\mathrm{V}}\approx 2.6$ for $z>2$, with a tail up to $A_{\mathrm{V}}\approx 4$ that is absent at lower $z$. Such a steep mass dependence is also reflected by the fraction of obscured star-formation in sub-mm detected galaxies \citep{whitaker17b, dunlop17, mclure18}. Of course, $A_{\mathrm{V}}$ has been inferred by many as part of SED fitting analysis for large samples of galaxies, but has not often been shown explicitly as a function of stellar mass and redshift. Only \citet{martis16} explicitly show this information, finding qualitatively similar trends: a higher $A_{\mathrm{V}}$ for more massive galaxies, and an increase with redshift from $A_{\mathrm{V}}\approx 0.8$ at $z<1$ to $A_{\mathrm{V}}\approx 1.5$ at $z>2$ \citep[also see][]{marchesini14}. But in quantitative terms, the mass- and redshift dependence are now much more pronounced and the $z\approx 2$ population extends to much redder colors and higher $A_{\rm V}$.

        Approximately 90\% of $M_\star>10^{11}~{\text{M}_\odot}$ star-forming galaxies at $1.5<z<2.5$ have $A_{\mathrm{V}}>2$, and $\approx 50\%$ have $A_{\mathrm{V}}>3$. Spectroscopic evidence for such high $A_{\mathrm{V}}$ was recently shown by \citet{cooper25, maheson25} and, likewise, some sub-mm detected galaxies are now shown to have equally high attenuation \citep{liu25}. 
        This stark evolution with redshift was under-illuminated in the past, because of the aforementioned underestimation of rest-frame $V-J$ colors of the reddest objects in most catalogs. To our knowledge, only \citet{martis19} found a tail of rest-frame $V-J>2$ colors and $A_{\mathrm{V}}>3$ in a near-IR selected sample at $z\approx 2-3$ that is absent at lower $z$. 
        We speculate that the reason for the decline in $A_{\mathrm{V}}$ with cosmic time is at least in part the consequence of the declining gas fraction. These massive galaxies likely have approximately solar metallicities at all $z$ probed here, suggesting that the dust-to-gas ratio may be similar across cosmic time. In addition, a change in dust geometry may play a role: where we see thin dust lanes in massive galaxies in the present-day Universe, the work by \citet{gebek25} suggests that the entire central region of a massive, star-forming galaxy is heavily obscured at $z>1.5$.
        It should be kept in mind that $A_{\mathrm{V}}$ estimates can be quite uncertain on a galaxy-by-galaxy basis \citep{pacifici23}. As a test, for our sub-sample of galaxies with $V-J>2.5$ we inferred $A_{\mathrm{V}}$ with a different SED modeling code \citep[\textsc{bagpipes},][]{carnall18}, allowing attenuation and metallicity to vary freely up to $A_{\rm V}=5$ \citep{calzetti00} and $Z=2.5Z_\odot$, and a double powerlaw star-formation history.  We find very comparable $A_{\mathrm{V}}$ values compared to \textsc{eazy}: 0.1 mag redder in the median, and 0.2 mag scatter, consistent with our uncertainties.   
        The increase in $A_{\mathrm{V}}$ with redshift does not lead to redder $U-V$ colors (Fig. \ref{fig:Global_UV-VJ_zbins_ccAv}), and sub-mm detected galaxies are only slightly redder than average, with large scatter. A high $A_{\mathrm{V}}$ with a relatively unreddened $U-V$ color requires a gray attenuation curve in the UV-optical. As shown recently by \citet{gebek25}, a plausible explanation is that not only young stars ($<100$~Myr), but all stars younger than $1$~Gyr -- a large fraction or even the majority of all stars at $z\approx 2$ -- are attenuated by a birth cloud-like component \citep{charlot00}.
        Note that in our lowest stellar mass bin $A_{\mathrm{V}}$ values remain small. It increases with redshift, but only slightly, from $A_{\mathrm{V}}\approx 0.3$ at $z<1$ to $A_{\mathrm{V}}\approx 0.4$ at $z>2$ (this, despite the bluer colors at higher $z$).

\section{Conclusion} \label{sec:conclusion}

Although it has been abundantly clear for decades that there exists a population of red galaxies at $z>1$, JWST/NIRCam presents a new opportunity to describe the distribution of rest-frame UV-optical-to-near-IR colors and $A_{\mathrm{V}}$ values for the general galaxy population up to $z=2.5$ and down to low stellar masses (here, $M_\star=10^9~M_\odot$). The rest-frame $U-V$-stellar mass diagram is familiar, and shows a distinct red sequence of passive galaxies and a population of star-forming galaxies that are blue at low mass and redder at higher mass. The rest-frame $V-J$-stellar mass distribution is remarkably different: the red sequence is more accurately described as a blue sequence: at $M_\star \gtrsim 3\times 10^{10}~M_\odot$, star-forming galaxies are much redder than quiescent galaxies, particularly at $z>1.5$. The red colors are explained by large $A_{\mathrm{V}}$ values. Nearly all massive star-forming galaxies at $z>1.5$ have $A_{\mathrm{V}}>2$. The photometric redshifts for the particular subset of extremely attenuated galaxies has markedly improved thanks for the precise NIRCam photometry, as shown by a comparison with spectroscopic redshift measurements, which explains why the strong redshift evolution in $V-J$ and $A_{\rm V}$ went largely unnoticed so far.

The overall goal of this paper has been to showcase the accurate and precise colors that the HST-JWST tandem affords, and provide a reference for the redshift evolution of rest-frame UV-optical-near-IR colors and $A_{\mathrm{V}}$ values as a function of stellar mass.
In a follow-up paper we show that the high $A_{\mathrm{V}}$ values are associated with strong radial color gradients, improving our understanding of the strong wavelength dependence of measured galaxy sizes and its interpretation.

A comprehensive spectroscopic survey of high-$A_{\mathrm{V}}$ galaxies at $z>1$, essentially a targeted extension of the results on sub-samples of other surveys that currently exist \citep{cooper25, maheson25}, is needed to understand the physical conditions in the centers of massive galaxies with large dust columns and high star-formation activity. Ground-based spectroscopic surveys are unavoidably biased toward the less-attenuated galaxies, not exceeding $A_{\mathrm{V}}\approx 1$ \citep[e.g.,][]{tacchella18, cullen18, lorenz24}. Furthermore, since the proportion of high-$A_{\mathrm{V}}$ is still rapidly increasing from $z=1.5$ to $z=2.5$, an extension to higher redshift is needed to determine when attenuation peaked in cosmic time. This may be linked to the population of $z\gtrsim 3$ optically dark, dust-obscured galaxies at $z\gtrsim 3$ \citep{franco18, barrufet23}. A redshift extension also serves to connect the narrative of massive, dusty galaxies at intermediate redshifts with the surprising presence of dusty galaxies at even higher redshifts  \citep[$z>5$; e.g.,][]{endsley23, akins23, barro24, shapley25, narayanan25, martis25}.

\begin{acknowledgements}
      MM acknowledges the financial support of the Flemish Fund for Scientific Research (FWO-Vlaanderen), research project G030319N.
      (Some of) The data products presented herein were retrieved from the Dawn JWST Archive (DJA). DJA is an initiative of the Cosmic Dawn Center (DAWN), which is funded by the Danish National Research Foundation under grant DNRF140. JSD acknowledges the support of the Royal Society via the award of a Research Professorship.
\end{acknowledgements}

\bibliographystyle{aa} % style aa.bst
%\bibliography{aanda}%, mypapers_arjen}

\end{document}